\documentclass[11pt]{article}

\usepackage{amsfonts,amsmath,amssymb}
\usepackage{enumerate}
\usepackage{hyperref}
\usepackage{bbm}
\usepackage{nicefrac}
\usepackage[all]{xy}
\usepackage{graphicx}
\usepackage{bm}

\usepackage{upgreek}

\usepackage{booktabs}

\newcommand{\be}{\begin{equation}}
\newcommand{\ee}{\end{equation}}

\newcommand{\bea}{\begin{eqnarray}}
\newcommand{\eea}{\end{eqnarray}}

\newcommand{\bes}{\begin{subequations}}
\newcommand{\ees}{\end{subequations}}

\newcommand{\cN}{{\cal N}}
\newcommand{\cA}{{\cal A}}

\def\ft#1#2{{\textstyle{{\scriptstyle #1}\over {\scriptstyle #2}}}}

\def\sst#1{{\scriptscriptstyle #1}}

\def\0{{\sst{(0)}}}
\def\1{{\sst{(1)}}}
\def\2{{\sst{(2)}}}
\def\3{{\sst{(3)}}}
\def\4{{\sst{(4)}}}
\def\5{{\sst{(5)}}}
\def\6{{\sst{(6)}}}
\def\7{{\sst{(7)}}}
\def\8{{\sst{(8)}}}

\def\cA{{{\cal A}}}
\def\cB{{{\cal B}}}
\def\cC{{{\cal C}}}
\def\cH{{{\cal H}}}
\def\cV{{{\cal V}}}
\def\cM{{{\cal M}}}

\def\tcA{{{\tilde{\cal A}}}}

\allowdisplaybreaks  

\usepackage{multirow}
\usepackage{rotating}

\allowdisplaybreaks

\let\OLDthebibliography\thebibliography
\renewcommand\thebibliography[1]{
  \OLDthebibliography{#1}
  \setlength{\parskip}{3.5pt}
}

\begin{document}

\begin{flushright}
CPHT-RR055.1215
\end{flushright}

   \begin{center}
   \baselineskip=16pt
   \begin{Large}\textbf{
The complete $D=11$ embedding of SO(8) supergravity}
   \end{Large}

\vspace{20pt}

\begin{large}
Oscar Varela
\end{large}

\vspace{10pt}

	\begin{small}	
          
   {\it Center for the Fundamental Laws of Nature,
	Harvard University, Cambridge, MA 02138, USA } \\

\vspace{5pt}
	
  {\it Centre de Physique Th\'eorique, Ecole Polytechnique, CNRS, 
	91128 Palaiseau Cedex, France } 
		
\vspace{5pt}

	ovarela@physics.harvard.edu
			
	\end{small}

\vskip 15pt

\end{center}

\begin{quote}

The truncation formulae of $D=11$ supergravity on $S^7$ to $D=4$ ${\cal N} =8$ SO(8)-gauged supergravity are completed to include the full non-linear dependence of the $D=11$ three-form potential $\hat A_\3$ on the $D=4$ fields, and their consistency is shown. The full embedding into $\hat A_\3$ is naturally expressed in terms of a restricted version, still ${\cal N} =8$ but only SL(8)--covariant, of the $D=4$ tensor hierarchy. The redundancies introduced by this approach are removed at the level of the field strength $\hat F_\4$ by exploiting $D=4$ duality relations. Finally, new expressions for the full consistent truncation formulae are given that are, for the first time, explicit in all $D=4$ fields.

\end{quote}


\vskip 5pt

\noindent {\bf Introduction}. Overwhelming evidence and partial proofs lend support to the expectation that $D=11$ supergravity \cite{Cremmer:1978km} admits a consistent Kaluza-Klein (KK) truncation on the seven-sphere to the purely electric SO(8) gauging \cite{deWit:1982ig} of maximal, $\cN=8$, supergravity in four dimensions \cite{Cremmer:1979up}. The complete consistent embedding of the four-dimensional metric $ds_4^2$, SO(8) gauge fields $\cA^{IJ} = \cA^{[IJ]}$, $I = 1, \ldots, 8$,  and $\textrm{E}_{7(7)}/\textrm{SU}(8)$ scalars into the eleven-dimensional metric is given by the classic de Wit-Nicolai result \cite{deWit:1986iy}
\begin{eqnarray} \label{KKFieldsIntro}
d\hat{s}_{11}^2 &=&  \Delta^{-1} \, ds_4^2  \, + g_{mn} \, Dy^m Dy^n  \; . 
\end{eqnarray}
Here, $y^m$, $m=1, \ldots, 7$, are the $S^7$ angles, $Dy^m \equiv  dy^m + \tfrac12 \, g \, K_{IJ}^m \, \cA^{IJ} $, with $g$ the $D=4$ gauge coupling constant, and $K_{IJ}^m$ are the $S^7$ Killing vectors. A formula for the combined dependence of the warp factor $\Delta$ and the {\it inverse} internal metric $g^{mn}$ on the $S^7$ angles and the $D=4$ scalars has been known for a long time \cite{deWit:1984nz,deWit:1986iy}. Recently, a similar formula has been given for the internal components $A_{mnp}$ of the $D=11$ three-form potential $\hat A_\3$ \cite{deWit:2013ija}.

These formulae define implicitly the dependence of the warp factor $\Delta$, of the internal {\it direct} metric $g_{mn}$ and of $A_{mnp}$ on the $D=4$ scalars: it is not obvious how to solve in full generality for these quantities directly from those equations. Only for smaller sectors of the $D=4$ theory, for which scalar parametrisations can be found explicitly, it does become possible to determine those quantities. In this way, $D=11$ solutions have been constructed from $D=4$ ones, thus strongly supporting the case for consistency. In any case, consistency has been manifestly established for some $\cN <8$ subtruncations \cite{Cvetic:1999xp,Cvetic:1999au} and, at the full $\cN=8$ level, for some sectors.  Indeed, the full $\cN=8$ embeddings into the metric, (\ref{KKFieldsIntro}), into the gravitino, and of the $D=4$ scalars into $A_{mnp}$ have been checked to hold consistently \cite{deWit:1986iy,deWit:2013ija}. An $\cN=8$ consistency proof, that excludes the three-forms, has also been given in \cite{Hohm:2014qga} from exceptional field theory \cite{Hohm:2013pua}. See also \cite{Lee:2014mla} for a similarly E$_{7(7)}$--covariant approach to consistency.


Up to the implicit dependence of  the $D=11$ metric (\ref{KKFieldsIntro}) on the $D=4$ scalars, the embedding (\ref{KKFieldsIntro}) is exact to all orders in the $D=4$ fields. In particular, the quadratic ({\it i.e.} linear in terms of the $D=11$ vielbein) dependence of the metric (\ref{KKFieldsIntro}) on the $D=4$ electric SO(8) vectors $\cA^{IJ}$ is exact. In contrast, much less is known about the full non-linear embedding of the $D=4$ fields into the three-form $\hat A_\3$ and its field strength $\hat F_\4$. Besides the non-linear dependence of $\hat A_\3$ on the $D=4$ scalars known implicitly \cite{deWit:2013ija}, a dependence of the former on the $D=4$ magnetic vectors $\tcA_{IJ}$ has been established at the linear level \cite{deWit:2013ija}. However, as we will show, this linear dependence is not exact in this case: $\hat A_\3$ contains, among others, non-linear couplings between the electric and magnetic $D=4$ vectors. 

In this article, the exact dependence of the $D=11$ three-form potential on the fields of $D=4$ $\cN=8$ supergravity is computed, and the consistency of the corresponding embedding formulae is shown. This completes the proof of  \cite{deWit:1986iy,deWit:2013ija} of the truncation of $D=11$ supergravity on $S^7$, at the level of the $\cN=8$ supersymmetry variations. We will focus on the purely electric SO(8) gauging \cite{deWit:1982ig} --an origin in conventional $D=11$ supergravity \cite{Cremmer:1978km} of its dyonic counterpart \cite{Dall'Agata:2012bb} has been ruled out \cite{deWit:2013ija,Lee:2015xga}. In order to find the full $D=11$ embedding, we will use new technology based on the $D=4$ tensor \cite{deWit:2007mt,deWit:2008ta} and duality \cite{Bergshoeff:2009ph} hierarchies. Similar techniques have been recently employed for the consistent embedding \cite{Guarino:2015jca,Guarino:2015vca} of dyonic ISO(7) supergravity \cite{Guarino:2015qaa} in massive IIA.

\vspace{10pt}

\noindent {\bf $D=11$ supergravity with only $\textrm{SO}(1,3)$ symmetry manifest}. In conventions in which the bosonic Lagrangian of $D=11$ supergravity \cite{Cremmer:1978km} reads
\begin{eqnarray}
{\cal L}_{11} \!\! &=& \!\! \hat R\,  \hat{\textrm{vol}}_{11} - 
\ft12 \, \hat F_\4 \wedge  {\hat *\hat F_\4}   -\ft16 \, \hat A_\3 \wedge \hat F_\4  \wedge \hat F_\4 \; ,\label{11DLagrangian}
\end{eqnarray}
%
%
the supersymmety transformations of the bosonic fields are
\begin{equation} \label{11DsusyVars}
 \delta   \hat{e}_M{}^A  = \tfrac{1}{4} \, \bar{\hat{\epsilon}} \, \hat{\Gamma}^A \,  \hat{\psi}_M \; , \qquad
 \delta   \hat{A}_{MNP} = - \tfrac{3}{4}  \,  \bar{\hat{\epsilon}} \, \hat{\Gamma}_{[MN} \,  \hat{\psi}_{P]}    \; ,
\end{equation}
where $M, N, \ldots = 0 , 1, \ldots , 10$ are world indices and $A, B, \ldots = 0 , 1, \ldots , 10$ are (only in (\ref{11DsusyVars}), not below) tangent-space indices. Following \cite{deWit:1986mz}, we consider a splitting of the $D=11$ local Lorentz symmetry as 
$
\textrm{SO}(1,10) \rightarrow \textrm{SO}(1,3) \times \textrm{SO}(7)
$, 
under which the eleven-dimensional coordinates and bosonic fields split as $(x^\mu, y^m)$, $\mu = 0 , 1, 2, 3$, $m = 1, \ldots , 7$ and
\begin{eqnarray} \label{KKFields}
d\hat{s}_{11}^2 &=&  \Delta^{-1} \, ds_4^2  \, + g_{mn} \big( dy^m + B^m \big)  \big( dy^n + B^n \big) \; , \nonumber \\[8pt]
\hat A_\3 &=&  \tfrac16 A_{\mu\nu\rho} \, dx^{\mu} \wedge dx^{\nu} \wedge dx^{\rho}  + \tfrac12  A_{\mu \nu m } \, dx^\mu \wedge dx^\nu \wedge \big( dy^m + B^m \big) \nonumber \\
&&  + \,  \tfrac12 A_{\mu mn}  \, dx^{\mu}   \wedge \big( dy^m + B^m \big) \wedge \big( dy^n + B^n \big) \nonumber \\
&& + \,  \tfrac16 A_{mnp} \big( dy^m + B^m \big) \wedge \big( dy^n + B^n \big) \wedge \big( dy^p + B^p \big) \; . 
\end{eqnarray}
Here,
$
\Delta^2 \equiv (\det \, g_{mn} ) /  ( \det \, \mathring g_{mn} )
$,
with $\mathring{g}_{mn}$ an arbitrary background metric. Straightforwardly promoting SO(7) in the splitting above to SL(7), the decompositions (\ref{KKFields}) give rise to $\textrm{SO}(1,3) \times \textrm{SL}(7)$--covariant $D=11$ bosonic fields. In the fermion sector, this $4+7$ split naturally leads to $\textrm{SO}(1,3) \times \textrm{SU}(8)$ covariant fermions: gravitini $\psi_\mu^A$, $\mu = 0 ,\ldots, 3$, $A=1, \ldots, 8$, coming from the `external' components of the fully-fledged SO$(1,10)$-covariant gravitino $\hat \psi_M$, and supersymmetry parameters $\epsilon^A$, in the $\bar{\bm{8}}$ of SU(8); and spin 1/2 fermions $\psi_a^A$, $a=1, \ldots, 7$, from the `internal' components of $\hat \psi_M$. It proves convenient to repack the latter into a trispinor 
$
\chi^{ABC} = \tfrac{3i}{\sqrt{2}} \,  (\Gamma^a C^{-1})^{[AB} \psi_a^{C]}
$ 
\cite{Cremmer:1979up}, which sits in the $\overline{\bm{56}}$ of SU(8). Here, $\Gamma_a$ and $C$ are the Dirac matrices and charge conjugation matrix in seven Euclidean dimensions. See \cite{Cremmer:1979up,deWit:1986mz} for further details on the fermion splitting and \cite{Guarino:2015vca} for our conventions. 

The $D=11$ supersymmetry transformations (\ref{11DsusyVars}) can now be expressed in terms of these fields. We find that the following further non-linear field redefinitions, 
\begin{eqnarray} \label{redeffields}
& C_\mu{}^{m8} \equiv  B_\mu{}^m \; , \qquad 
\tilde C_{\mu \, m n} \equiv  A_{\mu mn}   \; , \qquad
 C_{\mu\nu \, m}{}^8 \equiv -A_{\mu\nu m} +  C_{[\mu}{}^{n8} \,  \tilde C_{\nu]nm}  \; , \nonumber  \\[5pt] 
& C_{\mu\nu \rho}{}^{88} \equiv  A_{\mu\nu \rho} - C_{[\mu}{}^{m8} \, C_\nu{}^{n8} \,  \tilde{C}_{\rho]mn} \; , 
\end{eqnarray}
are necessary to bring the resulting $D=11$ transformations into a form compatible with the canonical $D=4$ $\cN=8$ expressions. Only the two- and three-forms are redefined with additional quadratic and cubic vector contributions. The vectors go unredefined --only for notational homogeneity have new symbols introduced for them in (\ref{redeffields}). Similar redefinitions have been previoulsy considered in $D=11$ with a dual six-form formulation \cite{Godazgar:2013dma}, in type IIB \cite{Ciceri:2014wya,Baguet:2015sma} and in type IIA \cite{Guarino:2015vca}. Introducing also the `generalised vielbeine' \cite{deWit:1986mz,deWit:2013ija},
\begin{eqnarray} \label{GenVielbein}
V^{m8}{}_{AB} =  \tfrac{1}{4} \,  \Delta^{-\frac12} \, e_a{}^m (C\Gamma^a)_{AB} \; ,  \quad
\tilde{V}_{mn \, AB} =   \tfrac{1}{4} \,   \Delta^{-\frac12} \, e_m{}^a e_n{}^b (C\Gamma_{ab})_{AB} + V^{p8}{}_{AB} \, A_{pmn} \; ,
\end{eqnarray}
a long calculation shows that the supersymmetry variations (\ref{11DsusyVars}) give rise to the variations given in \cite{deWit:1986mz,deWit:2013ija} (see \cite{Guarino:2015vca} for our conventions) for the scalars,  vectors and vielbein $e_\mu{}^\alpha$, and to the following new transformations for the two- (see nevertheless \cite{Godazgar:2014nqa}) and three-forms:
{\setlength\arraycolsep{0pt}
\begin{eqnarray} 
\label{eq:susySL72form}
&& \delta C_{\mu \nu \, m }{}^8  =  \Big[ \tfrac23 \big( V^{n8}{}_{BC} \,  \tilde{V}_{mn}{}^{AC} +  \tilde{V}_{mn \, BC}{} \,  V^{n8}{}^{AC} \big) \,  \bar{\epsilon}_{A} \gamma_{[\mu} \psi_{\nu]}^B  \nonumber    \\
&&  \qquad\qquad\quad + \tfrac{\sqrt{2}}{3} \,  V^{n8}{}_{AB} \,  \tilde{V}_{mn \, CD}  \,  \bar{\epsilon}^{[A} \gamma_{\mu \nu} \chi^{BCD]} +\textrm{h.c.} \Big] - C_{[\mu}{}^{n8} \, \delta \tilde{C}_{\nu] mn} -  \tilde{C}_{[\mu| \, mn} \,  \delta  C_{|\nu]}{}^{n8} 
 \; , \nonumber   \nonumber  \\[8pt]
\label{eq:susySL73form}
&& \delta C_{\mu \nu \rho}{}^{88} \,  =   \Big[ \tfrac{4i}{7}  \, V^{m8}{}_{BD} \, \big( V^{n8}{}^{DC} \,  \tilde{V}_{mn}{}_{AC} +  \tilde{V}_{mn}{}^{DC}{} \,  V^{n8}{}_{AC} \big) \,  \bar{\epsilon}^{A} \gamma_{[\mu \nu} \psi_{\rho]}^B \nonumber \\
&& \qquad\qquad\quad -i \tfrac{\sqrt{2}}{3} \, V^{m8 \, AE} \,   V^{n8}{}_{[EB|} \,  \tilde{V}_{mn \, |CD]}  \,  \bar{\epsilon}_A \gamma_{\mu \nu \rho} \chi^{BCD} +\textrm{h.c.} \Big]  \nonumber   \\
&&  \qquad\qquad\quad +3\, C_{[\mu\nu| \, m}{}^8 \,  \delta C_{|\rho]}{}^{m8} 
- C_{[\mu}{}^{m8} \,   \big( C_{\nu}{}^{n8} \, \delta \tilde{C}_{\rho] mn} +  \tilde{C}_{\nu| \, mn} \,  \delta  C_{|\rho]}{}^{n8} \big) \; .
\end{eqnarray}
}Here, $\gamma_\alpha$ are the SO$(1,3)$ gamma matrices and $\gamma_{\mu_1 \ldots \mu_p} \equiv e_{\mu_1}{}^{\alpha_1} \ldots e_{\mu_p}{}^{\alpha_p} \,\gamma_{\alpha_1 \ldots \alpha_p} $. These $D=11$ transformations are also compatible with $D=4$: they turn out to follow from branching the general E$_{7(7)}$--covariant $D=4$ $\cN=8$ transformations \cite{Cremmer:1979up,deWit:2007mt,Guarino:2015vca} and selecting appropriate SL(7) representations. Eqs.~(\ref{eq:susySL73form}) come out naturally written in the SL(8) basis --this is the origin of the scripts `8' that have been appended to some of the fields (\ref{redeffields}).

In summary, the $4+7$ splitting leads to $ \mathrm{SO}(1,3) \times  \mathrm{SL}(7) $--covariant $D=11$ bosons,
\begin{eqnarray} \label{eq:SL7fieldcontentCov}
\bm{1} &  \textrm{metric}:  &  \; ds_4^2 \, (x,y)  \; ,\nonumber \\
\bm{7}^\prime  + \bm{21} &  \textrm{generalised vielbeine}:  &   \; V^{m8}{}_{AB} (x,y)   \;, \;  \tilde{V}_{mn \, AB} (x,y)   \; , \nonumber \\
\bm{7}^\prime  + \bm{21} &  \textrm{vectors}:  &   \;  C_\mu{}^{m8} (x,y)  \;, \quad  \tilde{C}_{\mu \, mn} (x,y)    \; ,  \nonumber \\
\bm{7} &  \textrm{two-forms}:  &   \;  C_{\mu \nu \,  m}{}^8 (x,y)  \; , \nonumber \\
\bm{1} &  \textrm{three-form}:  &   \; C_{\mu \nu \rho}{}^{88} (x,y)   \; ,
\end{eqnarray}
and $ \mathrm{SO}(1,3) \times  \mathrm{SU}(8) $--covariant $D=11$ fermions $ \psi_\mu^A (x,y)$ and $\chi^{ABC} (x,y) $. The representations shown in (\ref{eq:SL7fieldcontentCov}) for the generalised vielbeine correspond to their SL(7) indices. Their (antisymmetric) indices $AB$ label the $\overline{\bm{28}}$ of SU(8), in agreement with their taking values along the antisymmetric combinations $C\Gamma_a$, $C\Gamma_{ab}$  of seven-dimensional gamma matrices.

\vspace{10pt}

\noindent {\bf A restricted $D=4$ $\cN=8$ duality hierarchy}. Recall that the conventional $D=4$ $\cN=8$ bosonic Lagrangian can be written in terms of the metric (through the Ricci tensor and the Hodge star); of the 28 electric SO(8) gauge fields $\cA^{IJ}$ (through their field strengths $\cH_{(2)}^{IJ}$); of the 70 scalars of the coset representative $\cV_\mathbb{M}{}^{ij} $, $\mathbb{M} = 1 , \ldots, 56$, $i=1 , \ldots , 8$,  of $\textrm{E}_{7(7)}/ \textrm{SU}(8)$ (through the square $\mathcal{M}_{\mathbb{MN}}= 2 \,  {\cal V_{( \mathbb{M}}}^{ij} \,  {\cal V}_{ \mathbb{N}) \, ij } $ and through the gauge kinetic matrices ${\cal I}_{[IJ][KL]} $ and $\mathcal{R}_{[IJ][KL]}$); and of the embedding tensor $\Theta_\mathbb{M}{}^\alpha$ (through the scalar potential $V$). Specifically \cite{deWit:1982ig,deWit:2007mt},
\begin{equation} 
\label{BosLag}
{\cal L} =  R \, \textrm{vol}_4  -\tfrac{1}{48} D {\cal M}_{\mathbb{M}\mathbb{N}} \wedge * D {\cal M}^{\mathbb{M}\mathbb{N}}  +\tfrac12 \, {\cal I}_{[IJ][KL]} \, \cH_{(2)}^{IJ} \wedge * \cH_{(2)}^{KL} + \tfrac12 \, \mathcal{R}_{[IJ][KL]} \, \cH_{(2)}^{IJ} \wedge  \cH_{(2)}^{KL} - V  \textrm{vol}_4 \; ,
\end{equation}
where $V$ is given by \cite{deWit:2007mt}
\begin{equation}
\label{N=8Pot}
V  =  \tfrac{1}{168} \, g^2 \,   {X_{\mathbb{MP}}}^{\mathbb{R}}  {X_{\mathbb{NQ}}}^{\mathbb{S}}  \mathcal{M}^{\mathbb{MN}}   \Big(  \mathcal{M}^{\mathbb{PQ}}  \mathcal{M}_{\mathbb{RS}}  +   7 \,   \delta^\mathbb{P}_\mathbb{S} \,  \delta^\mathbb{Q}_\mathbb{R}  \Big)  \; . 
\end{equation}
Here, ${\cal M}^{\mathbb{M} \mathbb{N}} = \Omega^{\mathbb{M} \mathbb{P}} \Omega^{\mathbb{N} \mathbb{Q}} {\cal M}_{\mathbb{P} \mathbb{Q}} $ is the inverse of ${\cal M}_{\mathbb{M} \mathbb{N}}$, with 
with $ \Omega^{\mathbb{M} \mathbb{N}}$ the Sp$(56,\mathbb{R})$-invariant matrix, and 
$
X_\mathbb{MN}{}^\mathbb{P} \equiv \Theta_\mathbb{M}{}^\alpha (t_\alpha)_\mathbb{N}{}^\mathbb{P} 
$
is the contraction of the embedding tensor $\Theta_\mathbb{M}{}^\alpha $ with the generators  $(t_\alpha)_\mathbb{N}{}^\mathbb{P} $ of E$_{7(7)}$. For the SO(8) gauging, only the 63 generators $(t_I{}^J)_\mathbb{N}{}^\mathbb{P} $ of the SL(8) subalgebra are relevant. Thus,
$
\Theta_\mathbb{M}{}^\alpha = (  \Theta_{[IJ]}{}^K{}_L  \; , \;   \Theta^{[IJ] K}{}_L )
$, with \cite{Dall'Agata:2012bb}
\begin{equation} \label{eq:SO8EmbTen}
\Theta_{[IJ]}{}^K{}_L = 2 \,  \delta^K_{[I} \theta_{J]L} \; , \qquad 
\Theta^{[IJ] K}{}_L =  2 \, \delta^{[I}_L \xi^{J]K} \; .
\end{equation}
Here, $\theta_{IJ} = \theta_{(IJ)}$ and $\xi^{IJ} = \xi^{(IJ)}$ generically lie in the $\bm{36}$ and $\bm{36}^\prime$ of SL(8), respectively. We will take 
$\theta_{IJ} = \delta_{IJ}$, $\xi^{IJ} = 0$,
corresponding to the purely electric SO(8) gauging of \cite{deWit:1982ig}. 

The field content that enters the conventional Lagrangian (\ref{BosLag}) can be enlarged to a so-called `tensor hierarchy' \cite{deWit:2008ta} that further includes, together with the magnetic vectors $\tcA_{IJ}$, two-, three- and four-form potentials in irreducible representations of E$_{7(7)}$.  The magnetic gauge fields and the two- and three-form potentials carry degrees of freedom, albeit not independent ones. Bringing the metric and scalars into the picture, their field strengths can be dualised into the dynamically independent fields that enter the Lagrangian (\ref{BosLag}). With these dualisations, the tensor hierarchy becomes promoted to a `duality hierarchy' \cite{Bergshoeff:2009ph}. By construction, the full E$_{7(7)}$-covariant tensor and duality hierarchies are closed under the Bianchi identities, the duality conditions and the $\cN=8$ supersymmetry variations. Here we will show that the following SL(8)--covariant  subsector of the hierarchy,
\begin{eqnarray} \label{eq:SL8fieldcontent4Dbis}
\bm{1} &  \textrm{metric}:  &  \; ds_4^2 (x) \nonumber \\
\bm{28}^\prime  + \bm{28}  &  \textrm{coset representatives}:  &   \; {\cal V}^{IJ \, ij} (x) \;, \;  \tilde{{\cal V}}_{IJ}{}^{ij} (x)   \; ,   \nonumber \\
\bm{28}^\prime + \bm{28}  &  \textrm{vectors}:  &   \; \cA^{IJ} (x)   \;, \quad   \tilde{\cA}_{IJ}  (x)   \; ,  \nonumber \\
\bm{63}  &  \textrm{two-forms}:  &   \; \cB_{I}{}^J  (x)   \; , \nonumber \\
\bm{36}^\prime &  \textrm{three-forms}:  &   \; \cC^{IJ}  (x) \; 
\end{eqnarray}
(which is still $\cN=8$ because all fermions $\psi_\mu^i $, $\chi^{ijk}$ are also retained), defines a closed subsector of the full tensor and duality hierarchies. The representations shown for the coset representative correspond to its SL(8) indices; its indices $ij$ label the $\overline{\bm{28}}$ of SU(8). All scalars and all (electric and magnetic) vectors are kept in (\ref{eq:SL8fieldcontent4Dbis}). However, only  two-form potentials $\cB_I{}^J$ (with $\cB_I{}^I = 0 $) in the adjoint of SL(8) and three-form potentials $\cC^{IJ} = \cC^{(IJ)}$ in the conjugate representation of the embedding tensor are retained.

Closure can be demonstrated by selecting the SL(8) representations shown in (\ref{eq:SL8fieldcontent4Dbis}) from the general  E$_{7(7)}$--covariant embedding-tensor-formalism results of \cite{deWit:2007mt,Bergshoeff:2009ph,Guarino:2015qaa}. For the supersymmetry variations of the two- and three-forms we find
{\setlength\arraycolsep{1pt}
\begin{eqnarray} \label{susytensors4dSL8}
\delta \cB_{\mu \nu \, J}{}^I & = &   \Big[ - \tfrac{2}{3} \big( {\cal V}^{IK}{}_{jk} \,  \tilde{{\cal V}}_{JK}{}^{ik}  +  \tilde{{\cal V}}_{JK \, jk}{} \,  {\cal V}^{IK}{}^{ik}   \big)  \bar{\epsilon}_{i} \gamma_{[\mu} \psi_{\nu]}^j  -  \tfrac{\sqrt{2}}{3} \,  {\cal V}^{IK}{}_{ij} \,  \tilde{{\cal V}}_{JK \, kl}   \,  \bar{\epsilon}^{[i} \gamma_{\mu \nu} \chi^{jkl]} +\textrm{h.c.} \Big]  \nonumber \\
&&  \quad  +  \big( \cA_{[\mu}^{IK} \, \delta \tilde{\cA}_{\nu] JK}  +  \tilde{\cA}_{[\mu| \, JK} \,  \delta  \cA_{|\nu]}{}^{IK}  \big) 
-\tfrac18 \,  \delta_J^I \,  (\textrm{trace}) \; , \nonumber \\[6pt]
 \label{susy3forms4dSL8}
\delta \cC_{\mu \nu \rho}{}^{IJ} & = &  \Big[ -  \tfrac{4i}{7} \,
 \cV^{K(I}{}_{jl} \big(  \cV^{J)L \, lk} \,  \tilde \cV_{KL \, ik}  + \tilde \cV_{ KL}{}^{ lk} \,  \cV^{J)L}{}_{ ik}   \big)   \, \bar{\epsilon}^{i}  \gamma_{[\mu \nu} \psi_{\rho]}^j  \nonumber \\
&& +i \tfrac{\sqrt{2}}{3}  \,
 \cV^{K(I|}{}^{hi} \,  \cV^{|J)L}{}_{[ij|} \,  \tilde \cV_{KL}{}_{|kl]}    \, \bar{\epsilon}_{h}  \gamma_{\mu \nu \rho } \chi^{jkl} \; + \textrm{h.c.} \Big] \nonumber \\
&& -3 \, \cB_{[\mu \nu| K}{}^{(I} \, \delta \cA^{J)K}_{|\rho]}  
+   \cA^{K(I}_{[\mu} \big(  \cA^{J)L}_\nu \,  \delta \tilde \cA_{\rho] KL}  + \tilde \cA_{\nu KL} \, \delta \cA^{J)L}_{\rho]}   \big)    \; .
\end{eqnarray}
}Please refer to \cite{Cremmer:1979up,deWit:2013ija} for the transformations of the fermions and the remaining fields in (\ref{eq:SL8fieldcontent4Dbis}) (and to \cite{Guarino:2015qaa} for those in our conventions). Together with the $\cN=8$ fermions, only bosonic fields in the restricted tensor hierarchy (\ref{eq:SL8fieldcontent4Dbis}) enter the r.h.s.~of these variations. The fermions transform in turn  into scalars and vector field strengths (see {\it e.g.} \cite{deWit:2007mt}), all of which were retained in (\ref{eq:SL8fieldcontent4Dbis}). Closure of the $\cN=8$ variations thus holds.

The $p$-form potentials, $p=1,2,3$, in (\ref{eq:SL8fieldcontent4Dbis}) can be checked to enjoy the Bianchi identities
\begin{eqnarray}
\label{eq:TruncBianchis}
& D {\cal H}_{\2}^{IJ} = 0 \; , \qquad
D {\tilde{\cal H}}_{\2 IJ} = - 2 \, g \, {\cH_{\3 [I}}^{K}\,\delta_{J]K}   \; , \qquad 
d {\cal H}_{\4}^{IJ} \equiv 0 \ , \nonumber \\[3pt] 
& D {\cal H}_{\3 I}{}^J = \cH_\2^{JK} \wedge \tilde{\cH}_{\2 IK}    -2 g\, \delta_{IK} \, \cH_\4^{JK} - \tfrac18 \, \delta_I^J \,  (\textrm{trace})   \ ,
\end{eqnarray}
which, indeed, also close among their field strengths. Using the embedding tensor formalism with (\ref{eq:SO8EmbTen}),  these field strengths can be computed to be
\begin{equation}
\label{eq:FormFieldStrengths}
\begin{array}{lll}
{\cal H}^{IJ}_\2 &=& d \cA^{IJ} - g \,  \delta_{KL} \, \cA^{IK} \wedge \cA^{LJ} \ , \\[10pt]
\tilde{{\cal H}}_{\2 IJ}  &=&  d \tilde{\cA}_{IJ} + g \,  \delta_{K[I} \, \cA^{KL}  \wedge \tilde{\cA}_{J]L}  +2 g \, \delta_{K[I} \, \cB_{J]}{}^K  \  ,  \\[10pt] 
{\cal H}_{\3 I}{}^J &=&  D \cB_I{}^J  + \tfrac12  \cA^{JK} \wedge d \tilde{\cA}_{IK}  + \tfrac12  \tilde{\cA}_{IK} \wedge d \cA^{JK}     - \tfrac 12 g \, \delta_{KL} \, \cA^{JK} \wedge \cA^{LM} \wedge \tilde{\cA}_{IM}   \\[4pt]
&&  + \tfrac 16 g \, \delta_{IK} \, \cA^{JL} \wedge \cA^{KM} \wedge \tilde{\cA}_{LM}     - 2 g \, \delta_{IK} \, \cC^{JK}  - \tfrac18 \, \delta_I^J \,  (\textrm{trace}) ,  \\[10pt] 
{\cal H}_{\4}^{IJ} & = & D \cC^{IJ}     - {\cal H}_\2^{K(I} \wedge \cB_K{}^{J)}   - \tfrac16  \cA^{K(I} \wedge \tilde{\cA}_{KL} \wedge d \cA^{J)L}   
+ \tfrac16  \cA^{IK} \wedge  \cA^{JL}  \wedge d \tilde{\cA}_{KL}  \\[4pt]
&&   -\tfrac16 g\, \delta_{KL} \, \cA^{K(I} \wedge {\cA}^{J)M} \wedge \cA^{LN} \wedge \tilde{\cA}_{MN} \; . \\[-10pt]
\end{array}
\end{equation}
Finally, the following closed duality relations can be calculated for these field strengths:
{\setlength\arraycolsep{2pt}
\begin{eqnarray}
\label{H2IJDuality}
\tilde{\cH}_{\2 IJ} & = & \tfrac{1}{2} {\cal I}_{[IJ][KL]} \, *{\cal H}_\2^{KL} + \tfrac{1}{2} {\cal R}_{[IJ][KL]} \, {\cal H}_\2^{KL}  \; ,  \qquad
\label{H3IJDuality}
\cH_{\3 I}{}^J =   \tfrac{1}{12}  (t_I{}^J)_{\mathbb{M}}{}^\mathbb{P}  \, \cM_{\mathbb{N}\mathbb{P}} *D \cM^{\mathbb{M} \mathbb{N}} \ , \nonumber  \\[10pt]
\label{H4Duality}
\cH_{\4}^{IJ} &= &  \tfrac{1}{84}  {X_{\mathbb{NQ}}}^{\mathbb{S}} \,(t_K{}^{(I|})_{\mathbb{P}}{}^{\mathbb{R}}    \mathcal{M}^{|J)K \, \mathbb{N}} \,  \big(  \mathcal{M}^{\mathbb{PQ}}  \mathcal{M}_{\mathbb{RS}}  +   7 \,   \delta^\mathbb{P}_\mathbb{S} \,  \delta^\mathbb{Q}_\mathbb{R}  \big)  \textrm{vol}_4  \ . \,\, \qquad 
\end{eqnarray}
}It follows from (\ref{H4Duality}) that $\cH_\4^{IJ}$ and the scalar potential $V$, given in (\ref{N=8Pot}), are related through
\begin{eqnarray}
\label{H4/Potential2}
 g \, \delta_{IJ} \, \cH_{\4}^{IJ}  = - 2 \, V \, \textrm{vol}_4 \; .
\end{eqnarray}

The combination of the duality relations with the Bianchi identities of the full E$_{7(7)}$-covariant tensor hierarchy produces the electric vector and projections of the scalar equations of motion \cite{Bergshoeff:2009ph}. All vectors were retained in the restricted hierarchy (\ref{eq:SL8fieldcontent4Dbis}). Accordingly, 
(\ref{eq:TruncBianchis}), (\ref{H4Duality}) give rise to the equations of motion for $\cA^{IJ}$ that derive from the Lagrangian (\ref{BosLag}). The Bianchi identity for $\cH_{\3 I}{}^J$ in (\ref{eq:TruncBianchis}), in the $\bm{63}$ of SL(8), corresponds to the projection to the SL(8) generators of the scalar equations of motion. Evaluated for constant scalars and vectors, this equation thus provides a relation that must be satisfied at every critical point of the scalar potential $V$. Splitting this equation into representations of SO$(8)$, $\bm{63} \rightarrow \bm{35}_v + \bm{28}$, only the components in the $\bm{35}_v$ can be found to be non-vanishing:
\begin{eqnarray} \label{Extreme35}
\big( \cH_\4^{IJ}   - \tfrac18 \, \delta^{IJ} \,  \delta_{KL} \, \cH_\4^{KL} \big) |_{0}  = 0 \; . 
\end{eqnarray}
Here, $|_0$ denotes evaluation at a critical point and $\cH_\4^{IJ} $ stands for its duality relation (\ref{H4Duality}).

\vspace{10pt}
 
\noindent {\bf The complete $S^7$ truncation}. The complete truncation ans\"atze are naturally given in terms of the $D=4$ restricted tensor hierarchy (\ref{eq:SL8fieldcontent4Dbis}). The KK ans\"atze relate linearly the $D=11$ $ \mathrm{SO}(1,3) \times  \mathrm{SL}(7) $--covariant fields (\ref{eq:SL7fieldcontentCov}) to the $D=4$ SL(8)--covariant ones (\ref{eq:SL8fieldcontent4Dbis}). The gap is bridged by $S^7$ tensors in appropriate $\mathrm{SL} (7) \times \mathrm{SL} (8)$ (or, equivalently, $\mathrm{SO} (7) \times \mathrm{SO} (8)$) representations. These include either constants or combinations of the coordinates $\mu^I(y)$ that embed $S^7$ into $\mathbb{R}^8$,
$
\delta_{IJ} \mu^I \mu^I = 1
$, 
and their derivatives $\partial_m \mu^I$ with respect to the $S^7$ angles $y^m$. Further combinations include the Killing vectors $K^{m \, IJ} = 2g^{-2} \, \mathring{g}^{mn} \, \mu^{[I} \partial_m \mu^{J]} $ of the round, SO(8)--symmetric metric $\mathring{g}_{mn}$ on $S^7$, and their derivatives $K_{mn}{}^{IJ} =  4g^{-2} \, \partial_{[m} \mu^I \partial_{n]} \mu^J $. Some useful identities satisfied by these tensors are
\begin{equation} \label{eq:propsS7}
g^{-2} \,  \mathring{g}^{mn} \partial_m \mu^I \partial_n \mu^J = \delta^{IJ} - \mu^I \mu^J \; , \;
K^m_{IJ} \,  \partial_m \mu^K = 2 \mu_{[I} \delta^K_{J]} \; , \;
K^m_{IJ} K_{mn}^{KL} = 8 g^{-2}  \mu_{[I} \delta^{[K}_{J]} \partial_n \mu^{L]} .
\end{equation}

The ansatz for the metric,
$
ds_4^2 (x,y) = ds_4^2 (x)
$, 
vectors,
$
C_\mu{}^{m8} (x,y)  = \tfrac12 \, g \, K_{IJ}^m (y) \, \cA_\mu{}^{IJ} (x) 
$,
$
\tilde C_{\mu \, mn} (x,y)  = \tfrac14 \, K_{mn}^{IJ} (y) \, \tilde \cA_{\mu \, IJ} (x) 
 $, 
scalars 
$
V^{m8 \, AB} (x,y) = \tfrac12 \, g \, K_{IJ}^m (y) \, \eta_i^A (y) \,  \eta_j^B (y) \, {\cal V}^{IJ \, ij} (x)
$, 
$
V_{mn}{}^{AB} (x,y)  =  \tfrac14 \, K_{mn}^{IJ} (y)  \, \eta_i^A (y) \,  \eta_j^B (y) \, \tilde {\cal V}_{IJ}{}^{ij} (x)
$ 
and fermions has been given and proved to be consistent, at the level of the $\cN=8$ transformations, in \cite{deWit:1986iy,deWit:2013ija}. In the KK ans\"atze for the generalised vielbeine, $\eta^A_i$ are the $S^7$ Killing spinors and $D=4$-scalar-dependent SU(8) rotations have been omitted. See \cite{deWit:1986iy,Nicolai:2011cy} for further details. Here we will extend the consistency proof of \cite{deWit:1986iy,deWit:2013ija} to the two- and three--forms. We first propose the ans\"atze
 \begin{eqnarray} \label{KKTwoForms}
C_{\mu \nu \, m}{}^8  (x,y) = -g^{-1}  \, ( \mu_I \partial_m \mu^J) (y) \, \cB_{\mu \nu \, J}{}^I (x)  \; , \qquad 
 C_{\mu \nu \rho}{}^{88}  (x,y) = ( \mu_I \mu_J) (y) \, \cC_{\mu \nu \rho}{}^{IJ} (x)  \ .
 \end{eqnarray}
Consistency can now be verified: when (\ref{KKTwoForms}) are inserted into the $D=11$ supersymmetry variations (\ref{eq:susySL73form}), a calculation involving the identities (\ref{eq:propsS7}) shows that all $S^7$ dependence drops out and the $D=4$ variations (\ref{susy3forms4dSL8}) arise. Essentially, the closure of the $D=4$ hierarchy (\ref{eq:SL8fieldcontent4Dbis}) devolves into the consistency of the $D=11$ truncation. This extends the results of \cite{deWit:1986iy,Nicolai:2011cy} and concludes the proof of the consistency of the truncation of $D=11$ supergravity on $S^7$, at the level of the $\cN=8$ supersymmetry transformations.

Equipped with the above consistent ans\"atze, the full $D=11$ embedding of $D=4$ SO(8)-gauged supergravity can be found by reworking backwards all intermediate steps, back to (\ref{KKFields}). To find the embedding of the $D=4$ $p$-forms (\ref{eq:SL8fieldcontent4Dbis}), $p=1,2,3$, the fields $B_\mu{}^m \, , \ldots$ that enter (\ref{KKFields}) can be expressed, inverting (\ref{redeffields}), in terms of the tensor-hierarchy-compatible fields, $C_\mu{}^{m8} \, ,\ldots$ and then these in terms of the KK ans\"atze. To determine the embedding of the scalars, the KK ans\"atze for the generalised vielbeine above first needs to be brought to (\ref{GenVielbein}). Then, taking all possible products between (\ref{GenVielbein}) and its conjugate, and tracing over SU(8) indices, all the dependence on the Killing spinors and the (omitted) SU(8) scalar matrices drops out and one is left with the following three equations:
{\setlength\arraycolsep{4pt}
\begin{eqnarray} 
{\cal M}^{IJ \, KL} \,  K^m_{IJ} \, K^n_{KL} &=&  4 g^{-2}   \, \Delta^{-1} \,  g^{mn}  \; , \label{eq:internalmetric} \\[5pt]
{\cal M}^{IJ}{}_{KL}  \, K^m_{IJ}  \, K_{np}^{KL}  &=&  8 g^{-1}  \,  \Delta^{-1} \,   g^{mq} \,  A_{qnp}   \; , \label{eq:internalAmnp} \\[5pt]
 {\cal M}_{IJ \, KL} \, K_{mn}^{IJ} \, K_{pq}^{KL} & = &   16 \, \Delta^{-1}  \Big( 2 \, g_{m[p} g_{q]n }   +g^{rs}  A_{rmn}  A_{spq} \Big) \; . \label{eq:additionalEqKK}
\end{eqnarray}
}The l.h.s.~feature the three SL(8)--covariant blocks of the $D=4$ scalar matrix $\cM_\mathbb{MN}$ contracted with combinations of the $S^7$ Killing vectors and their derivatives. On the r.h.s., the warp factor $\Delta$, the inverse, $g^{mn}$, and direct, $g_{mn}$, internal metric and three-form $A_{mnp}$ appear. Eq.~(\ref{eq:internalmetric}) has long been known \cite{deWit:1984nz,deWit:1986iy}, eq.~(\ref{eq:internalAmnp}) was recently derived in \cite{deWit:2013ija}, and eq.~(\ref{eq:additionalEqKK}) appears to be new at least in this explicit form --of course, it follows from \cite{deWit:2013ija,Godazgar:2013dma}. 

Combining (\ref{eq:internalmetric}), (\ref{eq:internalAmnp}) with the new equation (\ref{eq:additionalEqKK}), new formulae can be found for $g_{mn}$, $A_{mnp}$ and $\Delta$ which, unlike (\ref{eq:internalmetric}), (\ref{eq:internalAmnp}) by themselves, do provide explicit and independent expressions for those quantities. Bringing all these results to (\ref{KKFields}), the full non-linear embedding of (the restricted tensor hierarchy (\ref{eq:SL8fieldcontent4Dbis}) of) SO(8)-gauged supergravity into $D=11$ supergravity is finally obtained:
{\setlength\arraycolsep{0pt}
\begin{eqnarray} \label{KKEmbedding}
&& d\hat{s}_{11}^2 =  \Delta^{-1} \, ds_4^2  \, +\tfrac{1}{12} \,  g^{-2} \, \Delta^2 \, (t_I{}^J )_\mathbb{M}{}^\mathbb{P} \, (t_K{}^L )_\mathbb{N}{}^\mathbb{Q} \, \cM^\mathbb{MN}   \, \cM_\mathbb{PQ}  \, \mu_J \mu_L  D \mu^I D\mu^K      \; , \nonumber \\[10pt]
&& \hat A_\3 \,= \mu_I \mu_J  \, \big( \cC^{IJ}   +\tfrac16 \cA^{IK} \wedge \cA^{JL} \wedge \tilde{\cA}_{KL}    \big) + \, g^{-1}  \, \big( \cB_J{}^I  +\tfrac12 \cA^{IK} \wedge \tilde{\cA}_{KJ}  \big)  \wedge \mu_I D\mu^J    
\nonumber \\[3pt]
&& \qquad \quad +\tfrac12 g^{-2}\, \tilde \cA_{IJ} \wedge D\mu^I \wedge D \mu^J 
+ A \; .
\end{eqnarray}
}Here,  
$ 
D\mu^I  \equiv d \mu^I -g \, \delta_{JK}  \, \cA^{IJ} \mu^K 
$, 
the internal three-form $A \equiv   \tfrac16 A_{mnp} \, Dy^m \wedge Dy^n \wedge Dy^p$ is
\begin{eqnarray} \label{intA}
A = -\tfrac{1}{72} \, g^{-3} \, \Delta^3 \, (t_I{}^J)_\mathbb{P}{}^\mathbb{R} \, X^\prime_\mathbb{MQ}{}^\mathbb{S} \, \delta_\mathbb{NT} \, \Omega^\mathbb{TU} \, \Theta_\mathbb{U}{}^K{}_L \, \cM^\mathbb{MN} \,  \cM^\mathbb{PQ} \,  \cM_\mathbb{RS} \, \mu_J \, D \mu^I \wedge D \mu_K \wedge D\mu^L  
\end{eqnarray}
and the warp factor takes on the particularly elegant expression
\begin{equation}
\label{GeneralWarping}
\Delta^{-3}=   \tfrac{1}{84} \,  {X^\prime_{\mathbb{MP}}}^{\mathbb{R}}  {X^\prime_{\mathbb{NQ}}}^{\mathbb{S}}  \mathcal{M}^{\mathbb{MN}}   \Big(  \mathcal{M}^{\mathbb{PQ}}  \mathcal{M}_{\mathbb{RS}}  +   7 \,   \delta^\mathbb{P}_\mathbb{S} \,  \delta^\mathbb{Q}_\mathbb{R}  \Big)  \; , 
\end{equation}
which paralells the $D=4$ $\cN=8$ scalar potential (\ref{N=8Pot}). Following \cite{Guarino:2015vca},
in (\ref{intA}), (\ref{GeneralWarping}),
$
X^\prime_\mathbb{MN}{}^\mathbb{P} \equiv \Theta^\prime_\mathbb{M}{}^\alpha (t_\alpha)_\mathbb{N}{}^\mathbb{P} 
$, 
has been defined in terms of a `primed embedding' tensor 
$
\Theta^\prime_\mathbb{M}{}^\alpha = (  \Theta^\prime_{[IJ]}{}^K{}_L  \; , \;   \Theta^{\prime [IJ] K}{}_L )
$, 
which is in turn defined, exactly as in (\ref{eq:SO8EmbTen}), in terms of tensors $\theta^\prime_{IJ}$ and $\xi^{\prime IJ}$ in the $\bm{36}$ and $\bm{36}^\prime$ of SL(8). The magnetic $\xi^{\prime IJ}$ must vanish and the electric $\theta^\prime_{IJ}$ is now $S^7$-valued:
$
\theta^\prime_{IJ} = \mu_I \mu_J$, $\xi^{\prime IJ} \equiv 0$. 

Although the KK ans\"atze are linear in the $D=4$ $p$-forms, $p=1,2,3$, of the restricted tensor hierarchy (\ref{eq:SL8fieldcontent4Dbis}), $\hat A_\3$ develops a nonlinear dependence on the gauge fields by the redefinitions (\ref{redeffields}). In any case, it is interesting to consider the embedding also at the level of the field strength,  $\hat F_\4 = d \hat A_\3$. Differentiating (\ref{KKEmbedding}), a long calculation yields
\begin{equation} \label{eq:11DFS}
\hat F_\4 = \mu_I \mu_J  \, {\cal H}_\4^{IJ} + g^{-1} \,   {\cal H}_{\3 \, J}{}^I \wedge \mu_I D\mu^J + \tfrac12 \, g^{-2}  \, \tilde{{\cal H}}_{\2 IJ} \wedge D\mu^I \wedge D\mu^J+ dA  \; .
\end{equation}
Here, $ {\cal H}_\4^{IJ}$, $ {\cal H}_{\3  J}{}^I $ and $\tilde{{\cal H}}_{\2 IJ}$ turn out to coincide with the field strengths (\ref{eq:FormFieldStrengths}) of the restricted $D=4$ $\cN=8$ tensor hierarchy (\ref{eq:SL8fieldcontent4Dbis}). The fact that these $D=4$ field strengths can now be reproduced from an eleven-dimensional calculation provides a self-consistency check on our formalism. Moreover, (\ref{eq:11DFS}) can be used to establish the consistency of the truncation to the SO(8) gauging at the level of the Bianchi identities: some further calculation indeed shows that $d \hat F_\4 = 0$ devolves into the $D=4$ $\cN=8$ Bianchi identities (\ref{eq:TruncBianchis}). By the discussion below (\ref{H4/Potential2}), this automatically shows as well the consistency of the truncation at the level of the equations of motion of the SO(8) vectors and SL(8)/SO(8) scalars.

More interestingly, the $D=4$ duality hierarchy can be employed to eliminate the redundant degrees of freedom that are introduced in the $D=11$ embedding by the tensor hierarchy fields (\ref{eq:SL8fieldcontent4Dbis}). Trading the $D=4$ field strengths by their duality relations (\ref{H4Duality}) and now computing $dA$ explicitly from (\ref{intA}), the four-form field strength (\ref{eq:11DFS}) becomes
{\setlength\arraycolsep{0pt}
\begin{eqnarray} \label{eq:11DFSDualityHierarchy}
\hat F_\4 &=& \;\; U \, \textrm{vol}_4 + \tfrac{1}{12} \, g^{-1} \,   (t_I{}^J)_{\mathbb{M}}{}^\mathbb{P}  \, \cM_{\mathbb{N}\mathbb{P}} *D \cM^{\mathbb{M} \mathbb{N}}  \wedge \mu_J D\mu^I
\nonumber \\[6pt]
&+& \tfrac14 \, g^{-2}  \, \Big(  {\cal I}_{[IJ][KL]} \, *{\cal H}_\2^{KL} + {\cal R}_{[IJ][KL]} \, {\cal H}_\2^{KL}    \Big)  \wedge D\mu^I \wedge D\mu^J 
\nonumber \\[6pt]
&-&  \tfrac{1}{24} \, g^{-2} \, \Delta^3 \, (t_I{}^J)_\mathbb{P}{}^\mathbb{R} \, X^\prime_\mathbb{MQ}{}^\mathbb{S} \, \delta_\mathbb{NT} \, \Omega^\mathbb{TU} \, \Theta_\mathbb{U}{}^K{}_L \, \cM^\mathbb{MN} \,  \cM^\mathbb{PQ} \,  \cM_\mathbb{RS} \, \delta_{KM} \mu_J  \mu_N \, \cH_\2^{N[I} \wedge D \mu^M \wedge D\mu^{L]}   
\nonumber \\[8pt]
&-&  \tfrac{1}{6048} \, g^{-3} \, \Delta^6 \, (t_I{}^J)_\mathbb{P}{}^\mathbb{R} \,  \delta_\mathbb{NT} \, \Omega^\mathbb{TU} \, \Theta_\mathbb{U}{}^K{}_L  \nonumber \\
&& \quad  \times D \Big( X^\prime_\mathbb{MQ}{}^\mathbb{S} \, X^\prime_\mathbb{VX}{}^\mathbb{Z} \, X^\prime_\mathbb{WY}{}^\mathbb{A} \, \cM^\mathbb{MN} \,  \cM^\mathbb{PQ} \,  \cM^\mathbb{VW} \,  \cM^\mathbb{XY} \,  \cM_\mathbb{RS}\,  \cM_\mathbb{ZA} \Big) \wedge  \mu_J \, D \mu^I \wedge D \mu_K \wedge D\mu^L
\nonumber \\[8pt]
&-& \tfrac{1}{72} \, g^{-3} \, \Delta^3 \, (t_I{}^J)_\mathbb{P}{}^\mathbb{R} \, X^\prime_\mathbb{MQ}{}^\mathbb{S} \, \delta_\mathbb{NT} \, \Omega^\mathbb{TU} \, \Theta_\mathbb{U}{}^K{}_L \, \cM^\mathbb{MN} \,  \cM^\mathbb{PQ} \,  \cM_\mathbb{RS} \, D \mu_J \wedge D \mu^I \wedge D \mu_K \wedge D\mu^L   ,
\end{eqnarray}
}with the Freund-Rubin term elegantly given in terms of primed ($S^7$-dependent) and regular $D=4$ embedding tensors by the following expression, parallel to (\ref{N=8Pot}) and (\ref{GeneralWarping}):
\begin{equation}
\label{U_generalRewrite}
U \, \textrm{vol}_4 \,  \equiv \,  \cH_{\4}^{IJ}  \, \mu_{I} \, \mu_{J}   \, =  - \tfrac{1}{84} \, g \,  {X^\prime_{\mathbb{MP}}}^{\mathbb{R}}  {X_{\mathbb{NQ}}}^{\mathbb{S}}  \mathcal{M}^{\mathbb{MN}}   \Big(  \mathcal{M}^{\mathbb{PQ}}  \mathcal{M}_{\mathbb{RS}}  +   7 \,   \delta^\mathbb{P}_\mathbb{S} \,  \delta^\mathbb{Q}_\mathbb{R}  \Big) \, \textrm{vol}_4 \ .
\end{equation}
Equation (\ref{eq:11DFSDualityHierarchy}) is the first complete and explicit expression we are aware of for the full non-linear embedding of the $D=4$ SO(8) supergravity fields into $\hat F_\4$. Previous discussions include \cite{deWit:1986iy,Nicolai:2011cy,Godazgar:2015qia}. In contrast to the embedding (\ref{KKEmbedding}) into $\hat A_\3$, the embedding (\ref{eq:11DFSDualityHierarchy}) into $\hat F_\4$ is expressed in terms of the conventional fields (the metric, scalars and electric vectors) that enter the $D=4$ $\cN=8$ Lagrangian (\ref{BosLag}), along with the embedding tensor.

An equivalent rewrite of the Freund-Rubin contribution in terms of the potential (\ref{N=8Pot}) and the field strengths (which can be thought to stand for their duals (\ref{H4Duality})) is
{\setlength\arraycolsep{1pt}
\begin{equation} \label{FRForms}
\cH_{\4}^{IJ}  \, \mu_{I} \, \mu_{J}  = -\tfrac{1}{4} \, g^{-1} \, V \,  \textrm{vol}_{4} - \tfrac{1}{16} \, g^{-1} \,  \cH_\2^{IJ} \wedge \tilde{\cH}_{\2 IJ}   - \tfrac{1}{2} \, g^{-1} \, \big(D\cH_{\3 I}{}^{J} -\cH_\2^{JK} \wedge \tilde{\cH}_{\2 IK}  \big) \, \mu^{I}  \mu_{J}   \ ,
\end{equation}
}as some manipulation of the Bianchi identities (\ref{eq:TruncBianchis}) and the relation (\ref{H4/Potential2}) shows. At a critical point of the scalar potential, $\cH_{\3 I}{}^{J} =  \cH_\2^{IJ}  = \tilde{\cH}_{\2 IJ}  =0$, and (\ref{FRForms}) reduces to 
$
\cH_{\4}^{IJ} |_0  \, \mu_{I} \, \mu_{J}  = - \tfrac{1}{4} \, g^{-1} \, V_0 \,  \textrm{vol}_{4} 
$, 
where $|_0$ and $V_0$ denote evaluation at a critical point. The r.h.s.~becomes  independent of the $S^7$ coordinates, and so does the l.h.s, as can be seen by contracting the critical point condition (\ref{Extreme35}) with $\mu^I\mu^J$ and using the $S^7$ relation $\delta^{IJ} \mu_I \mu_J = 1$.
The Freund-Rubin term thus becomes $S^7$-independent and proportional to the cosmological constant $V_0$, in agreement with \cite{Nicolai:2011cy,Godazgar:2015qia}. More generally, $D\cH_{\3 I}{}^{J}$ in (\ref{FRForms}) contains, upon dualisation and use of the scalar equations of motion, contributions from the derivatives of $V$. An analogue observation has been made in \cite{Godazgar:2015qia}.

\vspace{10pt}

\noindent {\bf Final comments}. The new explicit consistent embedding formulae (\ref{KKEmbedding})--(\ref{GeneralWarping}),  (\ref{eq:11DFSDualityHierarchy}), (\ref{U_generalRewrite}) have been written in a basis-independent way that should facilitate their use in further computations. It is also straightforward to generalise them to describe hyperboloid truncations to SO$(p,q)$ gaugings, with $p+q=8$. The formulae are written with fundamental E$_{7(7)}$ indices (although only SL(8) adjoint indices). However, the attempt to turn on magnetic couplings $\xi^{IJ}$ \cite{Dall'Agata:2012bb} runs into the difficulties found in \cite{deWit:2013ija}. In any case, it would be interesting to explore the interplay of the present approach, based on the $D=4$ tensor \cite{deWit:2008ta} and duality \cite{Bergshoeff:2009ph} hierarchies, with the manifestly E$_{7(7)}$--covariant formalisms of \cite{Godazgar:2013dma,Lee:2014mla,Hohm:2014qga}.

\vspace{10pt}

\noindent {\bf Acknowledgements}. Collaboration with Adolfo Guarino on related projects is greatly appreciated. This work was supported by the Marie Curie fellowship PIOF-GA-2012-328798 and partially by the Fundamental Laws Initiative at Harvard. Hospitality from AEI, Potsdam, during write up is gratefully acknowledged.

\end{document}